\documentclass[10pt,a4paper,reqno]{amsart}
\usepackage{acronym}
\usepackage{amsfonts}
\usepackage{amsmath}
\usepackage{amssymb}
\usepackage{amsthm}
\usepackage{bm}
\usepackage{braket}
\usepackage{cite}
\usepackage{color}
\usepackage{geometry}
\usepackage{graphicx}
\usepackage{hyperref}
\usepackage{mathrsfs}
\usepackage{mathtools}
\usepackage{setspace}
\usepackage{stmaryrd}
\usepackage{subcaption}
\usepackage{tikz}

\newgeometry{top=1.5cm,bottom=1.5cm,outer=1.5cm,inner=1.5cm}

\newcommand{\bse}{\begin{subequations}}
\newcommand{\ese}{\end{subequations}}

\newtheorem{theorem}{Theorem}

\numberwithin{equation}{section}

\DeclareMathOperator{\tr}{tr}
\DeclareMathOperator{\diag}{diag}

\newcommand{\rT}{\mathrm{T}}
\newcommand{\rd}{\mathrm{d}}

\newcommand{\cN}{\mathcal{N}}

\newcommand{\bLd}{\mathbf{\Lambda}}
\newcommand{\tbLd}{{}^{t\!}\mathbf{\Lambda}}

\newcommand{\bO}{\mathbf{O}}

\newcommand{\bOa}{\mathbf{\Omega}}

\newcommand{\bC}{\mathbf{C}}

\newcommand{\bU}{\mathbf{U}}

\newcommand{\bc}{\mathbf{c}}
\newcommand{\tbc}{{}^{t\!}\mathbf{c}}
\newcommand{\bA}{\mathbf{A}}

\newcommand{\bI}{\mathbf{I}}

\newcommand{\ld}{\lambda}
\newcommand{\Ld}{\Lambda}
\newcommand{\tLd}{{}^{t\!}\Lambda}
\newcommand{\oa}{\omega}

\newcommand{\bSg}{\mathbf{\Sigma}}

\acrodef{2D}[2D]{two-dimensional}
\acrodef{2DTL}[2DTL]{two-dimensional Toda lattice}
\acrodef{3D}[3D]{three-dimensional}
\acrodef{ABS}[ABS]{Adler--Bobenko--Suris}
\acrodef{bDT}[bDT]{binary Darboux transform}
\acrodef{BT}[BT]{B\"acklund transform}
\acrodef{BSQ}[BSQ]{Boussinesq}
\acrodef{CAC}[CAC]{consistency-around-the-cube}
\acrodef{DT}[DT]{Darboux transform}
\acrodef{DL}[DL]{direct linearisation}
\acrodef{FX}[FX]{Fordy--Xenitidis}
\acrodef{GD}[GD]{Gel'fand--Dikii}
\acrodef{KP}[KP]{Kadomtsev--Petviashvili}
\acrodef{KdV}[KdV]{Korteweg--de Vries}
\acrodef{HM}[HM]{Hirota--Miwa}
\acrodef{MDC}[MDC]{multi-dimensional consistency}
\acrodef{NQC}[NQC]{Nijhoff--Quispel--Capel}
\acrodef{ODE}[ODE]{ordinary differential equation}
\acrodef{PDE}[PDE]{partial differential equation}
\acrodef{PDeltaE}[P$\Delta$E]{partial difference equation}
\acrodef{sG}[sG]{sine--Gordon}
\acrodef{YB}[YB]{Yang--Baxter}

\title[Direct linearisation approach to discrete integrable systems associated with $\mathbb{Z}_\mathcal{N}$ graded Lax pairs]
{Direct linearisation approach to discrete integrable systems associated with $\mathbb{Z}_\mathcal{N}$ graded Lax pairs}
\author{Wei Fu}
\address{School of Mathematical Sciences and Shanghai Key Laboratory of Pure Mathematics and Mathematical Practice \\
East China Normal University \\ 500 Dongchuan Road \\ Shanghai 200241 \\ People's Republic of China}

\begin{document}

\begin{abstract}
Fordy and Xenitidis [J. Phys. A: Math. Theor. \textbf{50} (2017) 165205] recently proposed a large class of discrete integrable systems
which include a number of novel integrable difference equations, from the perspective of $\mathbb{Z}_\mathcal{N}$ graded Lax pairs, without providing solutions.
In this paper, we establish the link between the Fordy--Xenitidis discrete systems in coprime case and linear integral equations in certain form,
which reveals solution structure of these equations.
The bilinear form of the Fordy--Xenitidis integrable difference equations is also presented together with the associated general tau function.
Furthermore, the solution structure explains the connections between the Fordy--Xenitidis novel models and the discrete Gel'fand--Dikii hierarchy.
\end{abstract}


\maketitle

\section{Introduction}\label{S:Intro}

The study of integrable difference equations is nowadays often considered as a core topic in the theory of integrable systems.
Technically, an integrable discrete equation may arise as the superposition formula for B\"acklund transforms of its corresponding continuous equation, see e.g. \cite{WE73}.
However, discrete systems in their own rights have remarkable mathematical significance due to the very rich algebraic structure behind them.
The key characteristic of an integrable \ac{PDE} is a Lie algebra, which completely determines the algebraic structure of the equation, see e.g. \cite{DS85,JM83}.
Compared with the continuous theory, the integrability of a \ac{PDeltaE} is, instead, governed by its associated Lie group.
It is such a group structure that makes discrete models algebraically much more general than continuous ones.
Reflecting on the nonlinear difference equation itself, this leads to the phenomenon that
a suitable continuum limit of a \ac{PDeltaE} obtained from the \ac{BT} approach normally yields the whole hierarchy of its corresponding \ac{PDE}s \cite{WC87}.
From this viewpoint, one could argue that discrete equations are the master models among integrable systems,
and continuous equations in a sense are the `approximation' of them.
However, most of the mathematical methods for integrable systems are invented based on the continuous theory,
and some times are no longer valid to the study of discrete models.
Therefore, to completely understand the structure of discrete systems,
it is essential to develop classical methods and invent novel concepts for the discrete theory, see monograph \cite{HJN16}.

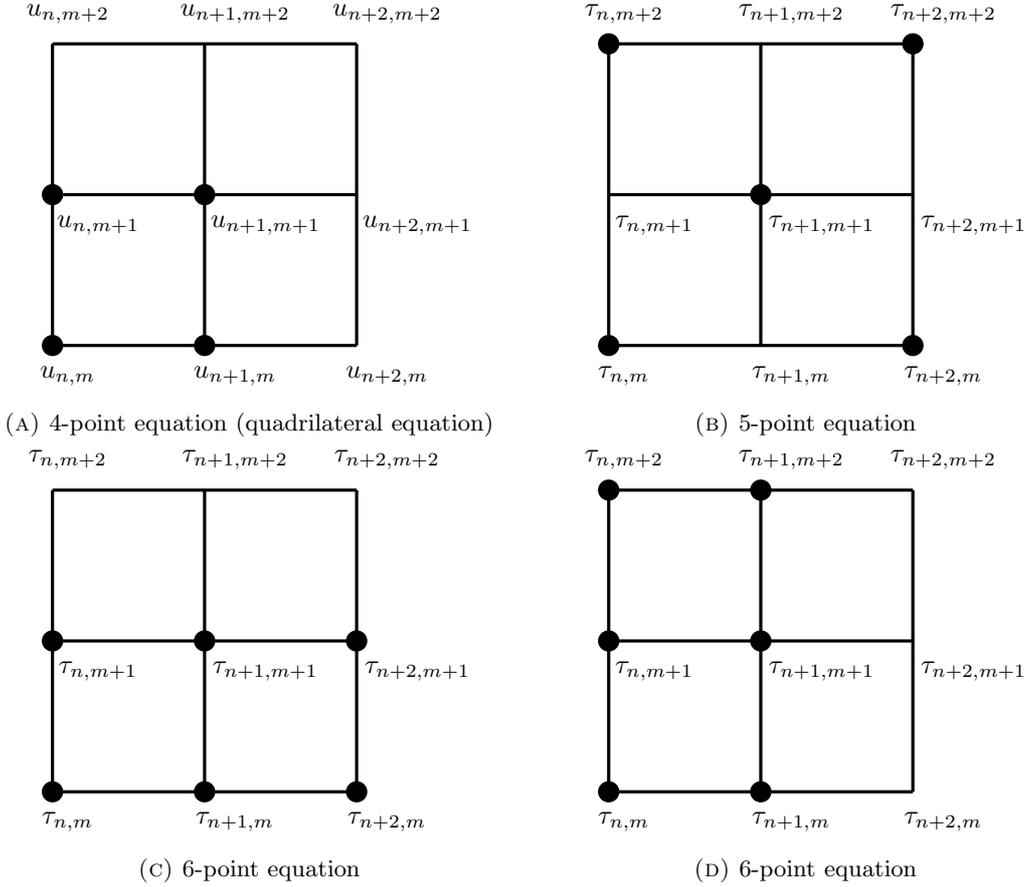
\begin{figure}
\centering
\begin{subfigure}[b]{0.4\textwidth}
\centering
\begin{tikzpicture}[scale=2]
 \coordinate (00) at (0, 0);
 \coordinate (10) at (1, 0);
 \coordinate (20) at (2, 0);
 \coordinate (01) at (0, 1);
 \coordinate (11) at (1, 1);
 \coordinate (21) at (2, 1);
 \coordinate (02) at (0, 2);
 \coordinate (12) at (1, 2);
 \coordinate (22) at (2, 2);
 \draw[very thick] (00) -- (20);
 \draw[very thick] (01) -- (21);
 \draw[very thick] (02) -- (22);
 \draw[very thick] (00) -- (02);
 \draw[very thick] (10) -- (12);
 \draw[very thick] (20) -- (22);
 \fill [black] (00) circle (2pt);
 \fill [black] (10) circle (2pt);
 \fill [black] (01) circle (2pt);
 \fill [black] (11) circle (2pt);
 \node at (0.1,-0.2) {$u_{n,m}$};
 \node at (1.2,-0.2) {$u_{n+1,m}$};
 \node at (2.2,-0.2) {$u_{n+2,m}$};
 \node at (0.3,0.8) {$u_{n,m+1}$};
 \node at (1.4,0.8) {$u_{n+1,m+1}$};
 \node at (2.4,0.8) {$u_{n+2,m+1}$};
 \node at (0.1,2.2) {$u_{n,m+2}$};
 \node at (1.2,2.2) {$u_{n+1,m+2}$};
 \node at (2.2,2.2) {$u_{n+2,m+2}$};
\end{tikzpicture}
\caption{4-point equation (quadrilateral equation)}
\label{F:4-Point}
\end{subfigure}
\begin{subfigure}[b]{0.4\textwidth}
\centering
\begin{tikzpicture}[scale=2]
 \coordinate (00) at (0, 0);
 \coordinate (10) at (1, 0);
 \coordinate (01) at (0, 1);
 \coordinate (11) at (1, 1);
 \draw[very thick] (00) -- (20);
 \draw[very thick] (01) -- (21);
 \draw[very thick] (02) -- (22);
 \draw[very thick] (00) -- (02);
 \draw[very thick] (10) -- (12);
 \draw[very thick] (20) -- (22);
 \fill [black] (00) circle (2pt);
 \fill [black] (20) circle (2pt);
 \fill [black] (11) circle (2pt);
 \fill [black] (02) circle (2pt);
 \fill [black] (22) circle (2pt);
 \node at (0.1,-0.2) {$\tau_{n,m}$};
 \node at (1.2,-0.2) {$\tau_{n+1,m}$};
 \node at (2.2,-0.2) {$\tau_{n+2,m}$};
 \node at (0.3,0.8) {$\tau_{n,m+1}$};
 \node at (1.4,0.8) {$\tau_{n+1,m+1}$};
 \node at (2.4,0.8) {$\tau_{n+2,m+1}$};
 \node at (0.1,2.2) {$\tau_{n,m+2}$};
 \node at (1.2,2.2) {$\tau_{n+1,m+2}$};
 \node at (2.2,2.2) {$\tau_{n+2,m+2}$};
\end{tikzpicture}
\caption{5-point equation}
\label{F:5-Point}
\end{subfigure}

\begin{subfigure}[b]{0.4\textwidth}
\centering
\begin{tikzpicture}[scale=2]
 \coordinate (00) at (0, 0);
 \coordinate (10) at (1, 0);
 \coordinate (20) at (2, 0);
 \coordinate (01) at (0, 1);
 \coordinate (11) at (1, 1);
 \coordinate (21) at (2, 1);
 \coordinate (02) at (0, 2);
 \coordinate (12) at (1, 2);
 \coordinate (22) at (2, 2);
 \draw[very thick] (00) -- (20);
 \draw[very thick] (01) -- (21);
 \draw[very thick] (02) -- (22);
 \draw[very thick] (00) -- (02);
 \draw[very thick] (10) -- (12);
 \draw[very thick] (20) -- (22);
 \fill [black] (00) circle (2pt);
 \fill [black] (10) circle (2pt);
 \fill [black] (20) circle (2pt);
 \fill [black] (01) circle (2pt);
 \fill [black] (11) circle (2pt);
 \fill [black] (21) circle (2pt);
 \node at (0.1,-0.2) {$\tau_{n,m}$};
 \node at (1.2,-0.2) {$\tau_{n+1,m}$};
 \node at (2.2,-0.2) {$\tau_{n+2,m}$};
 \node at (0.3,0.8) {$\tau_{n,m+1}$};
 \node at (1.4,0.8) {$\tau_{n+1,m+1}$};
 \node at (2.4,0.8) {$\tau_{n+2,m+1}$};
 \node at (0.1,2.2) {$\tau_{n,m+2}$};
 \node at (1.2,2.2) {$\tau_{n+1,m+2}$};
 \node at (2.2,2.2) {$\tau_{n+2,m+2}$};
\end{tikzpicture}
\caption{6-point equation}
\label{F:6-Point1}
\end{subfigure}
\begin{subfigure}[b]{0.4\textwidth}
\centering
\begin{tikzpicture}[scale=2]
 \coordinate (00) at (0, 0);
 \coordinate (10) at (1, 0);
 \coordinate (20) at (2, 0);
 \coordinate (01) at (0, 1);
 \coordinate (11) at (1, 1);
 \coordinate (21) at (2, 1);
 \coordinate (02) at (0, 2);
 \coordinate (12) at (1, 2);
 \coordinate (22) at (2, 2);
 \draw[very thick] (00) -- (20);
 \draw[very thick] (01) -- (21);
 \draw[very thick] (02) -- (22);
 \draw[very thick] (00) -- (02);
 \draw[very thick] (10) -- (12);
 \draw[very thick] (20) -- (22);
 \fill [black] (00) circle (2pt);
 \fill [black] (10) circle (2pt);
 \fill [black] (01) circle (2pt);
 \fill [black] (11) circle (2pt);
 \fill [black] (02) circle (2pt);
 \fill [black] (12) circle (2pt);
 \node at (0.1,-0.2) {$\tau_{n,m}$};
 \node at (1.2,-0.2) {$\tau_{n+1,m}$};
 \node at (2.2,-0.2) {$\tau_{n+2,m}$};
 \node at (0.3,0.8) {$\tau_{n,m+1}$};
 \node at (1.4,0.8) {$\tau_{n+1,m+1}$};
 \node at (2.4,0.8) {$\tau_{n+2,m+1}$};
 \node at (0.1,2.2) {$\tau_{n,m+2}$};
 \node at (1.2,2.2) {$\tau_{n+1,m+2}$};
 \node at (2.2,2.2) {$\tau_{n+2,m+2}$};
\end{tikzpicture}
\caption{6-point equation}
\label{F:6-Point2}
\end{subfigure}
\caption{Discrete KdV equation}
\label{F:dKdV}
\end{figure}

A milestone in the study of discrete integrable systems is
the \ac{ABS} classification \cite{ABS03} of scalar integrable quadrilateral equations (i.e. \ac{PDeltaE}s defined on four points).
In this classification, the most well-known model is the discrete \ac{KdV} equation (see Figure \ref{F:4-Point})
\bse\label{dKdV}
\begin{align}\label{dKdV:u}
 (p+q+u_{n,m}-u_{n+1,m+1})(p-q+u_{n,m+1}-u_{n+1,m})=p^2-q^2.
\end{align}
Here the subscripts $n$ and $m$ denote the lattice independent variables of the dependent variable $u$,
and $p$ and $q$ are their corresponding lattice parameters (we adopt such notations throughout the paper).
The discrete \ac{KdV} equation originally appeared as the superposition formula for the continuous potential \ac{KdV} equation,
see \cite{WE73}, and was later systematically studied together with the discrete modified \ac{KdV} equation
\begin{align}\label{dKdV:v}
 p(v_{n,m}v_{n,m+1}-v_{n+1,m}v_{n+1,m+1})=q(v_{n,m}v_{n+1,m}-v_{n,m+1}v_{n+1,m+1})
\end{align}
and the discrete Schwarzian \ac{KdV} equation (i.e. the cross-ratio equation)
\begin{align}\label{dKdV:z}
 \frac{(z_{n,m}-z_{n,m+1})(z_{n+1,m}-z_{n+1,m+1})}{(z_{n,m}-z_{n+1,m})(z_{n,m+1}-z_{n+1,m+1})}=\frac{p^2}{q^2}
\end{align}
by Nijhoff, Quispel and Capel et al., see e.g. the review paper \cite{NC95}.
Equations \eqref{dKdV:u}, \eqref{dKdV:v} and \eqref{dKdV:z} are the fundamental models in the \ac{ABS} classification,
and it was later shown in \cite{NAH09,ZZ13} that all the other equations in the \ac{ABS} list apart from Q4
(the master model in the \ac{ABS} classification whose algebraic structure is still not yet fully understood)
are all parameter extensions of these three equations.
There also exist integrable equations related to the \ac{ABS} equations which cannot be written in quadrilateral form.
In fact, the tau function $\tau$ associated with the above potentials $u$, $v$ and $z$ satisfies a 5-point equation (see Figure \ref{F:5-Point})
\begin{align}\label{dKdV:tau}
 (p-q)^2\tau_{n,m}\tau_{n+2,m+2}-(p+q)^2\tau_{n+2,m}\tau_{n,m+2}+4pq\tau_{n+1,m+1}^2=0.
\end{align}
There are also 6-point equations in the discrete KdV family as follows (see Figures \ref{F:6-Point1} and \ref{F:6-Point2}):
\begin{align}
 &(p+q)\tau_{n,m+1}\tau_{n+2,m}+(p-q)\tau_{n,m}\tau_{n+2,m+1}=2p\tau_{n+1,m}\tau_{n+1,m+1}, \label{dKdV:6-Point1} \\
 &(p+q)\tau_{n+1,m}\tau_{n,m+2}-(p-q)\tau_{n,m}\tau_{n+1,m+2}=2q\tau_{n,m+1}\tau_{n+1,m+1}. \label{dKdV:6-Point2}
\end{align}
\ese
Any of the above $\tau$-equations can be considered as the discrete bilinear \ac{KdV} equation,
since not only the continuum limit of each equation gives rise to the continuous bilinear \ac{KdV} equation,
but also the tau functions are exactly the same from the viewpoint of soliton-type solution,
describing structure of the discrete \ac{KdV} equation, see \cite{HJN16}.
However, we still needs to mark that from the aspect of initial value problem of lattice equations,
the 5-point equation and the 6-point equations are not equivalent.

Apart from quadrilateral equations, there also exist a large class of integrable difference equations defined on a $3\times 3$ stencil (i.e. 9-point equations which are second order in both $n$ and $m$), known as the discrete \ac{BSQ} family.
A typical model is the discrete \ac{BSQ} equation (see Figure \ref{F:dBSQ})
\bse\label{dBSQ}
\begin{align}\label{dBSQ:u}
 &\frac{p^3-q^3}{u_{n+2,m}-u_{n+1,m+1}}-\frac{p^3-q^3}{u_{n+1,m+1}-u_{n,m+2}} \nonumber \\
 &\qquad =u_{n+2,m+2}(u_{n+2,m+1}-u_{n+1,m+2})-u_{n+1,m}u_{n+2,m+1}+u_{n,m+1}u_{n+1,m+2}+u_{n,m}(u_{n+1,m}-u_{n,m+1}),
\end{align}
which was proposed by Nijhoff, Papageorgiou, Capel and Quispel \cite{NPCQ92} as a member in the discrete \ac{GD} hierarchy.
Besides \eqref{dBSQ:u}, there are also its Miura-related lattice equations, including the discrete modified \ac{BSQ} equation (see still \cite{NPCQ92})
\begin{align}\label{dBSQ:v}
 &\frac{v_{n,m}}{v_{n+1,m}}-\frac{v_{n,m}}{v_{n,m+1}}+\frac{v_{n+2,m+1}}{v_{n+2,m+2}}-\frac{v_{n+1,m+2}}{v_{n+2,m+2}} \nonumber \\
 &\qquad =\left(\frac{p^3v_{n+1,m+1}-q^3v_{n,m+2}}{v_{n,m+2}-v_{n+1,m+1}}\right)\frac{v_{n+1,m+2}}{v_{n,m+1}}
 -\left(\frac{p^3v_{n+2,m}-q^3v_{n+1,m+1}}{v_{n+1,m+1}-v_{n+2,m}}\right)\frac{v_{n+2,m+1}}{v_{n+1,m}},
\end{align}
and the discrete Schwarzian \ac{BSQ} equation \cite{Nij97} (a discrete equation involving multi-ratios)
\begin{align}\label{dBSQ:z}
 &\frac{(z_{n+2,m+2}-z_{n+1,m+2})(z_{n,m+2}-z_{n+1,m+1})(z_{n,m+1}-z_{n,m})}{(z_{n+2,m+2}-z_{n+2,m+1})(z_{n+2,m}-z_{n+1,m+1})(z_{n+1,m}-z_{n,m})} \nonumber \\
 &\qquad =\frac{p^3(z_{n+1,m+2}-z_{n,m+2})(z_{n+1,m+1}-z_{n,m+1})-q^3(z_{n+1,m+2}-z_{n+1,m+1})(z_{n,m+2}-z_{n,m+1})}
 {q^3(z_{n+2,m+1}-z_{n+2,m})(z_{n+1,m+1}-z_{n+1,m})-p^3(z_{n+2,m+1}-z_{n+1,m+1})(z_{n+2,m}-z_{n+1,m})}.
\end{align}
Very different from the \ac{KdV} case, namely the equations for the potential $\tau$ are written in bilinear form,
the tau function related to the determinant-type solution in the discrete BSQ family (see \cite{ZZN12,Fu17b} for exact formula) satisfies a \emph{trilinear} equation
\begin{align}\label{dBSQ:tau}
 &\left(p^2+pq+q^2\right)(\tau_{n+1,m}\tau_{n,m+2}\tau_{n+2,m+1}+\tau_{n,m+1}\tau_{n+2,m}\tau_{n+1,m+2}) \nonumber \\
 &\qquad =3p^2\tau_{n+1,m}\tau_{n+1,m+1}\tau_{n+1,m+2}+3q^2\tau_{n,m+1}\tau_{n+1,m+1}\tau_{n+2,m+1}-(p-q)^2\tau_{n,m}\tau_{n+1,m+1}\tau_{n+2,m+2}.
\end{align}
\ese
This may look surprising since the continuous \ac{BSQ} equation can be written as a bilinear equation.
However, if we take a suitable continuum limit of the 9-point equation \eqref{dBSQ:tau},
a common factor is generated in the procedure, and thus,
the continuum limit of the trilinear equation does give rise to the continuous bilinear \ac{BSQ} equation.
This gives us a strong hint that in the discrete theory, the bilinear structure may no longer be preserved.
The discrete \ac{BSQ}-type equations can alternatively be written as
three-component (effectively two-component) coupled systems of first-order \ac{PDeltaE}s.
A search for integrable discrete \ac{BSQ}-type systems was made by Hietarinta \cite{Hie11},
resulting in a remarkable classification of integrable multi-component \ac{BSQ}-type systems.
It was later proven that all Hietarinta's discrete \ac{BSQ}-type systems arise from the so-called extended discrete \ac{BSQ} systems \cite{ZZN12}.
\begin{figure}
\centering
\begin{tikzpicture}[scale=2]
 \coordinate (00) at (0, 0);
 \coordinate (10) at (1, 0);
 \coordinate (20) at (2, 0);
 \coordinate (01) at (0, 1);
 \coordinate (11) at (1, 1);
 \coordinate (21) at (2, 1);
 \coordinate (02) at (0, 2);
 \coordinate (12) at (1, 2);
 \coordinate (22) at (2, 2);
 \draw[very thick] (00) -- (20);
 \draw[very thick] (01) -- (21);
 \draw[very thick] (02) -- (22);
 \draw[very thick] (00) -- (02);
 \draw[very thick] (10) -- (12);
 \draw[very thick] (20) -- (22);
 \fill [black] (00) circle (2pt);
 \fill [black] (10) circle (2pt);
 \fill [black] (20) circle (2pt);
 \fill [black] (01) circle (2pt);
 \fill [black] (11) circle (2pt);
 \fill [black] (21) circle (2pt);
 \fill [black] (02) circle (2pt);
 \fill [black] (12) circle (2pt);
 \fill [black] (22) circle (2pt);
 \node at (0.1,-0.2) {$u_{n,m}$};
 \node at (1.2,-0.2) {$u_{n+1,m}$};
 \node at (2.2,-0.2) {$u_{n+2,m}$};
 \node at (0.3,0.8) {$u_{n,m+1}$};
 \node at (1.4,0.8) {$u_{n+1,m+1}$};
 \node at (2.4,0.8) {$u_{n+2,m+1}$};
 \node at (0.1,2.2) {$u_{n,m+2}$};
 \node at (1.2,2.2) {$u_{n+1,m+2}$};
 \node at (2.2,2.2) {$u_{n+2,m+2}$};
\end{tikzpicture}
\caption{Discrete BSQ equation (9-point equation)}
\label{F:dBSQ}
\end{figure}

The quadrilateral \ac{KdV}-type equations and 9-point \ac{BSQ} systems are integrable
in the sense of the so-called \ac{MDC} property
(i.e. a \ac{PDeltaE} can be consistently embedded into a higher-dimensional lattice space,
see e.g. Doliwa and Santini \cite{DS97}, Nijhoff and Walker \cite{NW01}, and also Bobenko and Suris \cite{BS02}).
The \ac{MDC} of the integrable quadrilateral equations is often presented as the \ac{CAC} phenomenon,
namely embedding a quadrilateral equation into a cube and showing the compatibility of equations defined on different faces,
see e.g. \cite{HJN16}.
However, the verification of the \ac{CAC} property is sometimes subtle.
The famous discrete \ac{sG} equation is a typical example --
one must consider simultaneously the discrete \ac{sG} and the discrete modified \ac{KdV} on a cube,
rather than verify its consistency directly;
to be more precise, we have the discrete {sG} on the top and bottom faces and the discrete modified \ac{KdV} on the four side faces on a single cube.
In other words, the discrete mKdV and sG flows commute.
Therefore, the key point behind the \ac{MDC} property is actually the existence of infinitely many commuting discrete flows of a difference equation,
which in turn results in the integrability (e.g. existence of a Lax pair, see \cite{Nij02,BS02}) of a difference equation.
More generally, an integrable discrete equation is compatible with not only its discrete symmetries,
but also its semi-discrete (semi-discrete hierarchy) and continuous symmetries (i.e. the continuous hierarchy);
in other words, discrete and continuous flow variables are on the same footing,
and the consistency is guaranteed by a unified solution structure behind all the equations, see \cite{Fu17b}.

The discrete \ac{KdV}-type and \ac{BSQ}-type equations are lower-order lattice models.
A natural problem would be to understand the integrability of higher-order (or higher-rank) \ac{PDeltaE}s.
However, studying higher-order difference equations directly is rather difficult
since even the 9-point \ac{PDeltaE}s already take complex forms.
A sensible way to solve such a problem is, instead, to consider integrable discrete multi-component systems,
which, in a sense, makes it possible for us to deal we equations in relatively lower orders,
though more components must be introduced as the price we have to pay.
Typical ways of constructing such systems include periodic reductions of discrete higher-dimensional models (see e.g. \cite{KNY02})
and the Lax approach based on ``big'' matrices (for instance, see \cite{PN96}), etc.

Very recently, Fordy and Xenitidis \cite{FX17a} generalised the early result
regarding the discrete-time Bogoyavlensky lattice equations in \cite{PN96},
and proposed a systematical scheme for integrable difference equations associated with $\mathbb{Z}_\cN$ graded discrete Lax pairs.
The key point in their construction is an $\cN\times\cN$ periodic matrix
\begin{align}\label{PeriodicMatrix}
 \bSg=
 \begin{pmatrix}
 0 & 1 & & & \\
 & 0 & 1 & \\
 & & \ddots & \ddots & \\
 & & & \ddots & 1 \\
 1 & & & & 0
 \end{pmatrix}_{\cN\times\cN},
\end{align}
which has very nice properties such as $\bSg^\cN=\bI$ and $\bSg^\rT=\bSg^{-1}$, where $\bI$ is the $\cN\times\cN$ unit matrix.
With the help of $\bSg$, $\mathbb{Z}_\cN$ graded discrete Lax pairs are generated by the $\cN\times\cN$ level $\alpha$ matrices
$\bA=\diag(a^{(0)},a^{(1)},\cdots,a^{(\cN-1)})\bSg^\alpha$ for $\alpha=0,1,\cdots,\cN-1$, i.e. the $\mathbb{Z}_\cN$ graded ring.
The benefit of such a construction is that all the integrable difference equations arising from such a framework
can be written as quadrilateral coupled systems and one thus successfully avoids dealing with higher-order terms.
Moreover, the integrability of these integrable difference equations also results in novel \ac{YB} maps \cite{FX17b}.

However, the exact solutions to the \ac{FX} models still needs to be clarified.
Our motivation is to investigate solution structure of these equations.
In this paper, we consider discrete equations in the equivalence class $[(\alpha,\alpha+1);(\beta,\beta+1)]$ within the \ac{FX} classification.
Integrable discrete equations in this case can mainly be expressed by either of the $\cN$-component discrete systems\footnote{
It is obvious that in equation \eqref{dFX:u} the lattice parameters $p$ and $q$ can be absorbed by introducing a simple point transform (i.e. a translation), but here we still reserve the lattice parameters in the equation.
This is because this treatment allows us to establish a more direct one-to-one correspondence between the system and its continuous analogue, and as a by-product, the procedure of taking continuum limit is also more natural.
}
\begin{align}\label{dFX:u}
 \frac{p+u_{n,m+1}^{(r+1+\alpha)}-u_{n+1,m+1}^{(r)}}{p+u_{n,m}^{(r+1+\alpha+\beta)}-u_{n+1,m}^{(r+\beta)}}
 =\frac{q+u_{n+1,m}^{(r+1+\beta)}-u_{n+1,m+1}^{(r)}}{q+u_{n,m}^{(r+1+\alpha+\beta)}-u_{n,m+1}^{(r+\alpha)}}, \quad
 u_{n,m}^{(r+\cN)}=u_{n,m}^{(r)},
\end{align}
and
\begin{align}\label{dFX:v}
 p\left(\frac{v_{n+1,m+1}^{(r)}}{v_{n,m+1}^{(r+\alpha)}}-\frac{v_{n+1,m}^{(r+1+\beta)}}{v_{n,m}^{(r+1+\alpha+\beta)}}\right)
 =q\left(\frac{v_{n+1,m+1}^{(r)}}{v_{n+1,m}^{(r+\beta)}}-\frac{v_{n,m+1}^{(r+1+\alpha)}}{v_{n,m}^{(r+1+\alpha+\beta)}}\right), \quad
 v_{n,m}^{(r+\cN)}= v_{n,m}^{(r)},
\end{align}
in which $r=0,1,\cdots,\cN-1$ for $\cN=2,3,\cdots$.
The variables $u$ and $v$ are referred to as the additive and quotient potentials, respectively,
describing the structure of unmodified and modified equations, respectively.
It was also pointed out in reference \cite{FX17a} that the additive potential $u$ and the quotient potential $v$ satisfy identities
\begin{align}\label{dFX:Identity}
 \prod_{r=0}^{\cN-1}\left(p+u_{n,m}^{(r+1+\alpha)}-u_{n+1,m}^{(r)}\right)=p^\cN, \quad
 \prod_{r=0}^{\cN-1}\left(q+u_{n,m}^{(r+1+\beta)}-u_{n,m+1}^{(r)}\right)=q^\cN
 \quad \hbox{and} \quad \prod_{r=0}^{\cN-1}v_{n,m}^{(r)}=1,
\end{align}
respectively.

These identities theoretically reduce the number of components in the discrete systems \eqref{dFX:u} and \eqref{dFX:v}.
For instance, when $(\alpha,\beta)=(0,0)$, $u_{n,m}^{(\cN-1)}$ and $v_{n,m}^{(\cN-1)}$ can be replaced by the other components through these identities,
which result in $(\cN-1)$-component systems
\begin{align*}
 &\frac{p+u_{n,m+1}^{(r+1)}-u_{n+1,m+1}^{(r)}}{p+u_{n,m}^{(r+1)}-u_{n+1,m}^{(r)}}
 =\frac{q+u_{n+1,m}^{(r+1)}-u_{n+1,m+1}^{(r)}}{q+u_{n,m}^{(r+1)}-u_{n,m+1}^{(r)}}, \quad
 r=0,1,\cdots,\cN-3, \\
 &\left(p+q+u_{n,m}^{(0)}-u_{n+1,m+1}^{(\cN-2)}\right)
 \left(p-q+u_{n,m+1}^{(\cN-2)}-u_{n+1,m}^{(\cN-2)}\right) \\
 &\qquad\qquad\qquad\qquad=\frac{p^\cN}{\prod_{r=0}^{\cN-3}\left(p+u_{n,m}^{(r+1)}-u_{n+1,m}^{(r)}\right)}
        -\frac{q^\cN}{\prod_{r=0}^{\cN-3}\left(q+u_{n,m}^{(r+1)}-u_{n,m+1}^{(r)}\right)}
\end{align*}
for the additive potential and
\begin{align*}
 &p\left(\frac{v_{n+1,m+1}^{(r)}}{v_{n,m+1}^{(r)}}-\frac{v_{n+1,m}^{(r+1)}}{v_{n,m}^{(r+1)}}\right)
 =q\left(\frac{v_{n+1,m+1}^{(r)}}{v_{n+1,m}^{(r)}}-\frac{v_{n,m+1}^{(r+1)}}{v_{n,m}^{(r+1)}}\right), \quad r=0,1,\cdots,\cN-3, \\
 &p\left(\frac{v_{n+1,m+1}^{(\cN-2)}}{v_{n,m+1}^{(\cN-2)}}
 -\frac{\prod_{r=0}^{\cN-2}v_{n,m}^{(r)}}{\prod_{r=0}^{\cN-2}v_{n+1,m}^{(r)}}\right)
 =q\left(\frac{v_{n+1,m+1}^{(\cN-2)}}{v_{n+1,m}^{(\cN-2)}}
 -\frac{\prod_{r=0}^{\cN-2}v_{n,m}^{(r)}}{\prod_{r=0}^{\cN-2}v_{n,m+1}^{(0)}}\right)
\end{align*}
for the quotient potential, i.e. the discrete \ac{GD} hierarchy (see \cite{NPCQ92}) and the discrete modified \ac{GD} hierarchy (see e.g. \cite{ALN12,Dol13});
similarly, in the $(\alpha,\beta)=(\cN-1,\cN-1)$ case, we have
\begin{align*}
 &\frac{p+u_{n,m+1}^{(0)}-u_{n+1,m+1}^{(0)}}{q+u_{n+1,m}^{(0)}-u_{n+1,m+1}^{(0)}}
 =\frac{p^\cN\prod_{r=0}^{\cN-2}\left(q+u_{n,m}^{(r)}-u_{n,m+1}^{(r)}\right)}
       {q^\cN\prod_{r=0}^{\cN-2}\left(p+u_{n,m}^{(r)}-u_{n+1,m}^{(r)}\right)}, \\
 &\frac{p+u_{n,m+1}^{(r)}-u_{n+1,m+1}^{(r)}}{q+u_{n+1,m}^{(r)}-u_{n+1,m+1}^{(r)}}
 =\frac{p+u_{n,m}^{(r-1)}-u_{n+1,m}^{(r-1)}}{q+u_{n,m}^{(r-1)}-u_{n,m+1}^{(r-1)}}, \quad
 r=1,2,\cdots,\cN-3,
\end{align*}
namely the discrete Schwarzian \ac{GD} hierarchy which was first proposed in \cite{Atk08}.
When $\cN=2$, the above three systems can be further rewritten as scalar equations \eqref{dKdV:v}, \eqref{dKdV:u} and \eqref{dKdV:z}, respectively;
while for $\cN=3$, they lead to the 9-point equations \eqref{dBSQ:v}, \eqref{dBSQ:u} and \eqref{dBSQ:z}, respectively.

Formulae \eqref{dFX:u} and \eqref{dFX:v} together with \eqref{dFX:Identity} not only cover the well-known discrete models mentioned above,
but also give rise to a huge number of new integrable multi-component discrete systems in case of $\cN\geq 3$, see \cite{FX17a} for more details.
For example, when $\cN=3$ and $(\alpha,\beta)=(1,1)$, the multi-component system of the quotient potential $v=v^{(0)}$ can reduce to a scalar integrable discrete equation
\begin{align}\label{dFX:9pt}
 \left(\frac{pv_{n+1,m+1}-qv_{n+2,m}}{pv_{n,m+2}-qv_{n+1,m+1}}\right)\left(\frac{p^2v_{n+1,m+1}-q^2v_{n,m+2}}{p^2v_{n+2,m}-q^2v_{n+1,m+1}}\right)=\frac{X_{n,m}X_{n,m+1}X_{n+1,m+1}}{Y_{n,m}Y_{n+1,m}Y_{n+1,m+1}}
\end{align}
where $X$ and $Y$ are given by
\begin{align*}
 X_{n,m}\doteq pv_{n,m}v_{n+1,m}v_{n+1,m+1}+q \quad \hbox{and} \quad Y_{n,m}\doteq qv_{n,m}v_{n,m+1}v_{n+1,m+1}+p,
\end{align*}
respectively, which is a novel 9-point lattice equation compared with \eqref{dBSQ:v}, \eqref{dBSQ:u} and \eqref{dBSQ:z}.

Both systems \eqref{dFX:u} and \eqref{dFX:v} are integrable in the sense of having their respective Lax pair.
Concretely, the compatibility condition of the $\cN$-component linear system (i.e. $\mathbb{Z}_\cN$ graded Lax pair)
\bse\label{dFX:Lax}
\begin{align}
 &\phi_{n+1,m}^{(r)}=\left(p+u_{n,m}^{(r+1+\alpha)}-u_{n+1,m}^{(r)}\right)\phi_{n,m}^{(r+\alpha)}+\phi_{n,m}^{(r+1+\alpha)}, \label{dFX:Lax1} \\
 &\phi_{n,m+1}^{(r)}=\left(q+u_{n,m}^{(r+1+\beta)}-u_{n,m+1}^{(r)}\right)\phi_{n,m}^{(r+\beta)}+\phi_{n,m}^{(r+1+\beta)}, \label{dFX:Lax2}
\end{align}
\ese
in which $\phi_{n,m}^{(r+\cN)}=k^\cN\phi_{n,m}^{(r)}$ and $r=0,1,\cdots,\cN-1$ for $\cN=2,3,\cdots$, gives rise to equation \eqref{dFX:u}.
Furthermore, the potentials $u$ and $v$ are connected via a difference transform \cite{FX17a}, i.e. a \ac{BT}, which is composed of
\begin{align}\label{dFX:MT}
 p+u_{n,m}^{(r+1+\alpha)}-u_{n+1,m}^{(r)}
 =p\frac{v_{n+1,m}^{(r)}}{v_{n,m}^{(r+\alpha)}} \quad \hbox{and} \quad
 q+u_{n,m}^{(r+1+\beta)}-u_{n,m+1}^{(r)}
 =q\frac{v_{n,m+1}^{(r)}}{v_{n,m}^{(r+\beta)}}.
\end{align}
Substituting $u$ with $v$ with the help of \eqref{dFX:MT} in \eqref{dFX:Lax}, we obtain the Lax pair for equation \eqref{dFX:v}.

The method we adopt is the so-called \ac{DL},
which was originally proposed by Fokas and Ablowitz in \cite{FA81} to solve the initial value problem of the \ac{KdV} equation.
It was later developed by Nijhoff, Quispel, Capel et al. to construct integrable discrete equations and their exact solutions, see e.g. \cite{NC95,NPCQ92}.
The advantage of the \ac{DL} is that it not only provides solutions to nonlinear equations, but also reveals the underlying structures of nonlinear integrable equations.
The main idea in the \ac{DL} approach is to associate a nonlinear equation with a linear integral equation.
In practice, by considering the nonlinearisation of a certain linear integral equation,
one can algebraically construct the corresponding nonlinear equation as well as its solutions and other integrability characteristics simultaneously.

The main results in this paper are listed below.
\begin{theorem}\label{T:DL}
Consider a linear integral equation
\begin{align}\label{dFX:Integral}
 \phi_{n,m}^{(r)}(k)
 +\sum_{j\in J}\int_{\Gamma_j}\rd\ld_j(l)\frac{\rho_{n,m}^{(r)}(k)\sigma_{n,m}^{(r)}(\oa^jl)}{k-\oa^jl}\phi_{n,m}^{(r)}(l)
 =\rho_{n,m}^{(r)}(k),
\end{align}
where $\rho$ and $\sigma$ are the plane wave factors are expressions of the discrete variables $n$, $m$ and $r$
(which are associated with their respective lattice parameters $p$, $q$ and $0$) as well as the spectral parameter $k$ (or $l$), given by
\begin{align}\label{dFX:PWF}
 \rho_{n,m}^{(r)}(k)=(k^\alpha(p+k))^{n}(k^\beta(q+k))^{m}k^r \quad \hbox{and} \quad
 \sigma_{n,m}^{(r)}(k')=(k'^\alpha(p+k'))^{-n}(k'^\beta(q+k'))^{-m}k'^{-r};
\end{align}
$\phi$ is the wave function depending on the discrete variables $n$, $m$, $r$ and also the spectral parameter;
$\rd\ld_j$ are the measures for the integrations on their respective contours $\Gamma_j$ (which are arbitrary and can be specified for particular classes of solutions);
$\oa\doteq\exp\{2\pi\mathfrak{i}/\cN\}$ with $\mathfrak{i}$ being the imaginary unit is a primitive $\cN$th root of unity;
and $J$ is a set defined as $J\doteq\{j|\hbox{$0<j<\cN$ are integers coprime to $\cN$}\}$.

For any solution $\phi$ to the linear integral equation \eqref{dFX:Integral}, the additive and quotient potentials given by
\begin{align}\label{dFX:uv}
 u_{n,m}^{(r)}\doteq\sum_{j\in J}\int_{\Gamma_j}\rd\ld_j(k)\phi_{n,m}^{(r)}(k)\sigma_{n,m}^{(r)}(\oa^jk) \quad \hbox{and} \quad
 v_{n,m}^{(r)}\doteq1+\sum_{j\in J}\int_{\Gamma_j}\rd\ld_j(k)\phi_{n,m}^{(r)}(k)\sigma_{n,m}^{(r)}(\oa^jk)\oa^{-j}k^{-1},
\end{align}
respectively, solve equations \eqref{dFX:u} and \eqref{dFX:v}, respectively, and $\phi$ solves the corresponding linear system \eqref{dFX:Lax}.
\end{theorem}

\begin{theorem}\label{T:BL}
Equations \eqref{dFX:u} and \eqref{dFX:v} are both linked with the bilinear system
\begin{align}\label{dFX:tau}
 p\left(\tau_{n,m}^{(r+1+\alpha+\beta)}\tau_{n+1,m+1}^{(r)}-\tau_{n,m+1}^{(r+\alpha)}\tau_{n+1,m}^{(r+1+\beta)}\right)
 =q\left(\tau_{n,m}^{(r+1+\alpha+\beta)}\tau_{n+1,m+1}^{(r)}-\tau_{n+1,m}^{(r+\beta)}\tau_{n,m+1}^{(r+1+\alpha)}\right),
\end{align}
in which $\tau_{n,m}^{(r+\cN)}=\tau_{n,m}^{(r)}$, and $r=0,1,\cdots,\cN-1$ for $\cN=2,3,\cdots$ through
\begin{align}\label{dFX:BLT}
 p+u_{n,m}^{(r+1)}-u_{n+1,m}^{(r)}=p\frac{\tau_{n,m}^{(r)}\tau_{n+1,m}^{(r+1)}}{\tau_{n,m}^{(r+1)}\tau_{n+1,m}^{(r)}}, \quad
 q+u_{n,m}^{(r+1)}-u_{n,m+1}^{(r)}=q\frac{\tau_{n,m}^{(r)}\tau_{n,m+1}^{(r+1)}}{\tau_{n,m}^{(r+1)}\tau_{n,m+1}^{(r)}} \quad \hbox{and} \quad
 v_{n,m}^{(r)}=\frac{\tau_{n,m}^{(r+1)}}{\tau_{n,m}^{(r)}}.
\end{align}
The bilinear system \eqref{dFX:tau} is solved by
\begin{align}\label{dFX:tauFunction}
 \tau_{n,m}^{(r)}=\sum_{i'=0}^{\infty}\frac{1}{i'!}\left(\sum_{i=0}^\infty\frac{(-1)^{i}}{i+1}\tr\left(\bOa\bC_{n,m}^{(r)}\right)^{i+1}\right)^{i'},
\end{align}
where $\bOa$ is an infinite matrix defined as $\bOa=\sum_{i=0}^\infty\tbLd^i\bO\bLd^{-i-1}$
with $\bLd$, $\tbLd$ and $\bO$ being infinite matrices having their respective $(i,j)$-entries
\begin{align*}
 \Ld(i,j)=\delta_{i+1,j}, \quad \tLd(i,j)=\delta_{i,j+1} \quad \hbox{and} \quad O(i,j)=\delta_{i,0}\delta_{0,j};
\end{align*}
the infinite matrix $\bC$ is defined as
\begin{align}\label{dFX:C}
 \bC_{n,m}^{(r)}=\sum_{j\in J}\int_{\Gamma_j}\rd\ld_j(k)\rho_{n,m}^{(r)}(k)\bc(k)\tbc(\oa^jk)\sigma_{n,m}^{(r)}(\oa^jk)
\end{align}
in which $\bc$ and $\tbc$ are an infinite-dimensional column vector and an infinite-dimensional row vector, respectively, defined as follows:
\begin{align*}
 \bc(k)=(\cdots,k^{-2},k^{-1},1,k,k^2,\cdots)^\rT, \quad \tbc(k')=(\cdots,k'^{-2},k'^{-1},1,k',k'^2,\cdots).
\end{align*}
\end{theorem}
Theorem \ref{T:DL} establishes the link between equations \eqref{dFX:u} and \eqref{dFX:v}, and the linear integral equation \eqref{dFX:Integral}.
This theoretically provides us with a way to solve the two discrete nonlinear systems.
Theorem \ref{T:BL} gives us not only a novel nonlinear form (i.e. the bilinear form) of the \ac{FX} equations, but also its solution.

\section{Proof of theorem \ref{T:DL}}\label{S:DL}

We consider a more general linear integral equation
\begin{align}\label{Integral}
 \Phi_{n,m}^{(h)}(k)+\sum_{j\in J}\int_{\Gamma_j}\rd\ld_j(l)\frac{\rho_{n,m}^{(h)}(k)\sigma_{n,m}^{(h)}(\oa^jl)}{k-\oa^jl}\Phi_{n,m}^{(h)}(l)=\rho_{n,m}^{(h)}(k)\bc(k),
\end{align}
in which the wave function $\Phi$ is an infinite column vector with $i$th component $\phi_{n,m}^{(h)}(i;k)$
as a function depending on the discrete variables $n$, $m$ and $h$ as well as the spectral variable $k$ (or $l$);
and the plane wave factors are given by
\begin{align}\label{dGD:PWF}
 \rho_{n,m}^{(h)}(k)=(p+k)^n(q+k)^mk^h \quad \hbox{and} \quad
 \sigma_{n,m}^{(h)}(k')=(p+k')^{-n}(q+k')^{-m}k'^{-h},
\end{align}
namely the $(\alpha,\beta)=(0,0)$ case of \eqref{dFX:PWF}.
Compared with \eqref{dFX:Integral}, \eqref{Integral} is a set of linear integral equations.
The reason why we start with such a linear integral equation is because it allows us to introduce the so-called infinite matrix formalism,
which will significantly simplify the algebraic construction of discrete nonlinear equations.
The idea of the whole proof is to link equation \eqref{Integral} with \eqref{dFX:u} and \eqref{dFX:v} for the $(\alpha,\beta)=(0,0)$,
and then by introducing the parameter deformation with respect to $\alpha$ and $\beta$, we complete the proof of theorem \ref{T:DL}.

We introduce a new quantity
\begin{align}\label{U}
 \bU_{n,m}^{(h)}\doteq\sum_{j\in J}\int_{\Gamma_j}\rd\ld_j(k)\Phi_{n,m}^{(h)}(k)\tbc(\oa^j k)\sigma_{n,m}^{(h)}(\oa^j k).
\end{align}
The quantity $\bU$ is an infinite matrix having its $(i,j)$-entry $U_{n,m}^{(h)}(i,j)$ only depending on the discrete variables $n$, $m$ and $h$,
and independent of the spectral parameter. It plays the role of nonlinearisation of the wave function $\Phi$.
On the other hand, following the definition of $\bOa$, we can verify that
\begin{align*}
 \tbc(\oa^j l)\bOa\bc(k)=\sum_{i=0}^\infty(\oa^j l)^{-i-1}k^i=\frac{1}{k}\sum_{i=0}^\infty\left(\frac{\oa^jl}{k}\right)^i
 =\frac{1}{k}\frac{1}{1-\oa^j l/k}=\frac{1}{k-\oa^jl}
\end{align*}
directly. Replacing $\frac{1}{k-\oa^jl}$ by $\tbc(\oa^j l)\bOa\bc(k)$ in the integral equation \eqref{Integral}, we have
\begin{align*}
 \rho_{n,m}^{(h)}(k)\bc(k)&=\Phi_{n,m}^{(h)}(k)+\sum_{j\in J}\int_{\Gamma_j}\rd\ld_j(l)\Phi_{n,m}^{(h)}(l)\frac{\rho_{n,m}^{(h)}(k)\sigma_{n,m}^{(h)}(\oa^jl)}{k-\oa^jl} \\
 &=\Phi_{n,m}^{(h)}(k)+\sum_{j\in J}\int_{\Gamma_j}\rd\ld_j(l)\Phi_{n,m}^{(h)}(l)\rho_{n,m}^{(h)}(k)\tbc(\oa^j l)\bOa\bc(k)\sigma_{n,m}^{(h)}(\oa^jl) \\
 &=\Phi_{n,m}^{(h)}(k)+\left(\sum_{j\in J}\int_{\Gamma_j}\rd\ld_j(l)\Phi_{n,m}^{(h)}(l)\sigma_{n,m}^{(h)}(\oa^jl)\tbc(\oa^j l)\right)\bOa\bc(k)\rho_{n,m}^{(h)}(k);
\end{align*}
in other words, equation \eqref{Integral} can be reformulated as a compact form
\begin{align}\label{Linear}
 \Phi_{n,m}^{(h)}(k)=\left(1-\bU_{n,m}^{(h)}\bOa\right)\bc(k)\rho_{n,m}^{(h)}(k).
\end{align}
Taking the definition of $\bU$ (i.e. \eqref{U}) into consideration and acting the operation
$\sum_{j\in J}\int_{\Gamma_j}\rd\ld_j(k)\cdot\tbc(\oa^j k)\sigma_{n,m}^{(h)}(\oa^j k)$ on the linear integral equation \eqref{Linear},
we immediately obtain its nonlinear version, i.e. a relation for the quantity $\bU$, as follows:
\begin{align}\label{Nonlinear}
 \bU_{n,m}^{(h)}=\left(1-\bU_{n,m}^{(h)}\bOa\right)\bC_{n,m}^{(h)}, \quad \hbox{which is equivalent to} \quad
 \bU_{n,m}^{(h)}=\bC_{n,m}^{(h)}\left(1+\bOa\bC_{n,m}^{(h)}\right)^{-1},
\end{align}
where $\bC$ is an infinite matrix defined as
\begin{align}\label{C}
 \bC_{n,m}^{(h)}\doteq\sum_{j\in J}\int_{\Gamma_j}\rd\ld_j(k)\rho_{n,m}^{(h)}(k)\bc(k)\tbc(\oa^jk)\sigma_{n,m}^{(h)}(\oa^jk),
\end{align}
which is actually the infinite matrix representation of the effective plane wave factor
\begin{align}\label{ePWF}
 \rho_{n,m}^{(h)}(k)\sigma_{n,m}^{(h)}(\oa^jk)=\left(\frac{p+k}{p+\oa^jk}\right)^n\left(\frac{q+k}{q+\oa^j k}\right)^m\left(\frac{1}{\oa^j}\right)^h.
\end{align}

We now algebraically construct equations \eqref{dFX:u} and \eqref{dFX:v} based on the infinite matrix $\bU$.
Since $\oa^j$ has the property $(\oa^j)^\cN=(\oa^\cN)^j=1$, we can easily prove from \eqref{ePWF} that the effective plane wave factor satisfies the periodic condition
$\rho_{n,m}^{(h+\cN)}(k)\sigma_{n,m}^{(h+\cN)}(\oa^jk)=\rho_{n,m}^{(h)}(k)\sigma_{n,m}^{(h)}(\oa^jk)$.
This immediately results in the fact that the infinite matrix $\bC$ obeys the periodic condition
\begin{align}\label{dGD:Period}
 \bC_{n,m}^{(h+\cN)}=\bC_{n,m}^{(h)}, \quad \hbox{and consequently} \quad \bU_{n,m}^{(h+\cN)}=\bU_{n,m}^{(h)},
\end{align}
as the dynamics of $\bU$ relies on $\bC$ explicitly according to \eqref{Nonlinear},
which implies that the discrete variable $h$ from now on can only be selected from a finite set,
namely it will play the role of labelling components in the resulting nonlinear equations.
Notice that
\begin{align*}
 \bC_{n+1,m}^{(h)}&=\sum_{j\in J}\int_{\Gamma_j}\rd\ld_j(k)\rho_{n,m}^{(h)}(k)(p+k)\bc(k)\tbc(\oa^jk)(p+\oa^jk)^{-1}\sigma_{n,m}^{(h)} \\
 &=\sum_{j\in J}\int_{\Gamma_j}\rd\ld_j(k)\rho_{n,m}^{(h)}(p+\bLd)\bc(k)\tbc(\oa^jk)(p+\tbLd)^{-1}\sigma_{n,m}^{(h)}=(p+\bLd)\bC_{n,m}^{(h)}(p+\tbLd)^{-1},
\end{align*}
as well as $\bOa\bLd-\tbLd\bOa=\bO$. By shifting equation \eqref{Nonlinear} with respect to $n$ by one unit, one obtains
\begin{align*}
 \bU_{n+1,m}^{(h)}&=\left(1-\bU_{n+1,m}^{(h)}\bOa\right)\bC_{n+1,m}^{(h)}=\left(1-\bU_{n+1,m}^{(h)}\bOa\right)(p+\bLd)\bC_{n,m}^{(h)}(p+\tbLd)^{-1} \\
 &=(p+\bLd)\bC_{n,m}^{(h)}(p+\tbLd)^{-1}-\bU_{n+1,m}^{(h)}\bOa(p+\bLd)\bC_{n,m}^{(h)}(p+\tbLd)^{-1} \\
 &=(p+\bLd)\bC_{n,m}^{(h)}(p+\tbLd)^{-1}-\bU_{n+1,m}^{(h)}\left[\bO+(p+\tbLd)\bOa\right]\bC_{n,m}^{(h)}(p+\tbLd)^{-1},
\end{align*}
which can equivalently be written as
\begin{align*}
 \bU_{n+1,m}^{(h)}(p+\tbLd)\left(1+\bOa\bC_{n,m}^{(h)}\right)=(p+\bLd)\bC_{n,m}^{(h)}-\bU_{n+1,m}^{(h)}\bO\bC_{n,m}^{(h)}.
\end{align*}
Multiplying this equation by $\left(1+\bOa\bC_{n,m}^{(h)}\right)^{-1}$ and noticing equation \eqref{Nonlinear}, we finally reach
\bse\label{dGD:UDyn}
\begin{align}
 \bU_{n+1,m}^{(h)}(p+\tbLd)=(p+\bLd)\bU_{n,m}^{(h)}-\bU_{n+1,m}^{(h)}\bO\bU_{n,m}^{(h)}. \label{dGD:UDyna}
\end{align}
Similarly, considering the shifts of $\bU$ with respect to $m$ and $h$ gives rise to the dynamics of $\bU$ as follows:
\begin{align}
 &\bU_{n,m+1}^{(h)}(q+\tbLd)=(q+\bLd)\bU_{n,m}^{(h)}-\bU_{n,m+1}^{(h)}\bO\bU_{n,m}^{(h)}, \label{dGD:UDynb} \\
 &\bU_{n,m}^{(h+1)}\tbLd=\bLd\bU_{n,m}^{(h)}-\bU_{n,m}^{(h+1)}\bO\bU_{n,m}^{(h)}. \label{dGD:UDync}
\end{align}
\ese

Equations listed in \eqref{dGD:UDyn} are the fundamental relations to algebraically construct the nonlinear equations.
The main idea of the construction is to focus on certain entries of $\bU$ and prove that these entries form their respective closed-form nonlinear equations,
which will later lead to equations \eqref{dFX:u} and \eqref{dFX:v}. For this purpose, we introduce new variables
\begin{align}\label{uv}
 u_{n,m}^{(h)}=U_{n,m}^{(h)}(0,0) \quad \hbox{and} \quad v_{n,m}^{(h)}=1+U_{n,m}^{(h)}(0,-1).
\end{align}
It is obvious to see that $u$ and $v$ satisfy
\begin{align}\label{dGD:uvPeriod}
 u_{n,m}^{(h+\cN)}=u_{n,m}^{(h)} \quad \hbox{and} \quad v_{n,m}^{(h+\cN)}=v_{n,m}^{(h)},
\end{align}
because they are effectively the entries of $\bU$ and the periodicity is preserved.
Taking the $(0,0)$-entry of \eqref{dGD:UDyn}, the dynamical equations in \eqref{dGD:UDyn} give rise to the following equations involving only $u$, $U(1,0)$ and $U(0,1)$:
\begin{align*}
 &p u_{n+1,m}^{(h)}+U_{n+1,m}^{(h)}(0,1)=pu_{n,m}^{(h)}+U_{n,m}^{(h)}(1,0)-u_{n+1,m}^{(h)}u_{n,m}^{(h)}, \\
 &q u_{n,m+1}^{(h)}+U_{n,m+1}^{(h)}(0,1)=qu_{n,m}^{(h)}+U_{n,m}^{(h)}(1,0)-u_{n,m+1}^{(h)}u_{n,m}^{(h)}, \\
 &U_{n,m}^{(h+1)}(0,1)=U_{n,m}^{(h)}(1,0)-u_{n,m}^{(h+1)}u_{n,m}^{(h)},
\end{align*}
We aim to eliminate $U(1,0)$ and $U(0,1)$ and derive a closed-form equation for only $u$.
In practice, by subtraction the entry $U(1,0)$ can be eliminated and we obtain
\bse
\begin{align}
 &U_{n+1,m}^{(h)}(0,1)-U_{n,m+1}^{(h)}(0,1)=p\left(u_{n,m}^{(h)}-u_{n+1,m}^{(h)}\right)-q\left(u_{n,m}^{(h)}-u_{n,m+1}^{(h)}\right)+u_{n,m}^{(h)}\left(u_{n,m+1}^{(h)}-u_{n+1,m}^{(h)}\right), \label{dGD:a} \\
 &U_{n,m+1}^{(h)}(0,1)-U_{n,m}^{(h+1)}(0,1)=q\left(u_{n,m}^{(h)}-u_{n,m+1}^{(h)}\right)+u_{n,m}^{(h)}\left(u_{n,m}^{(h+1)}-u_{n,m+1}^{(h)}\right), \label{dGD:b}\\
 &U_{n,m+1}^{(h)}(0,1)-U_{n+1,m}^{(h)}(0,1)=-p\left(u_{n,m}^{(h)}-u_{n+1,m}^{(h)}\right)+u_{n,m}^{(h)}\left(u_{n+1,m}^{(h)}-u_{n,m}^{(h+1)}\right). \label{dGD:c}
\end{align}
\ese
The summation $\eqref{dGD:a}_{n,m}^{(h+1)}+\eqref{dGD:b}_{n+1,m}^{(h)}+\eqref{dGD:c}_{n,m+1}^{(h)}$ can help to eliminate the entry $U(0,1)$,
and thus, we end up with a closed-form discrete equation expressed by only $u$, which is in the form of
\begin{align}
 \frac{p+u_{n,m+1}^{(h+1)}-u_{n+1,m+1}^{(h)}}{p+u_{n,m}^{(h+1)}-u_{n+1,m}^{(h)}}
 =\frac{q+u_{n+1,m}^{(h+1)}-u_{n+1,m+1}^{(h)}}{q+u_{n,m}^{(h+1)}-u_{n,m+1}^{(h)}}. \label{dGD:u}
\end{align}
This equation together with the periodic condition for $u$ given in \eqref{dGD:uvPeriod}, namely $u_{n,m}^{(h+\cN)}=u_{n,m}^{(h)}$,
forms the multi-component system \eqref{dFX:u} for $(\alpha,\beta)=(0,0)$, when $h$ is fixed from $0,1,\cdots,\cN-1$.
Next, we construct the closed-form equation for the potential $v$. The $(0,-1)$-entry of \eqref{dGD:UDyn} gives us the following:
\begin{align*}
 &p\left(v_{n+1,m}^{(h)}-v_{n,m}^{(h)}\right)+u_{n+1,m}^{(h)}v_{n,m}^{(h)}=U_{n,m}^{(h)}(1,-1), \\
 &q\left(v_{n,m+1}^{(h)}-v_{n,m}^{(h)}\right)+u_{n,m+1}^{(h)}v_{n,m}^{(h)}=U_{n,m}^{(h)}(1,-1), \\
 &u_{n+1,m}^{(h)}v_{n,m}^{(h)}=U_{n,m}^{(h)}(1,-1).
\end{align*}
Eliminating $U(1,-1)$ in these relations, the following difference equations involving $u$ and $v$ can be derived:
\bse\label{dGD:MT}
\begin{align}
 &p-q+u_{n,m+1}^{(h)}-u_{n+1,m}^{(h)}=p\frac{v_{n+1,m}^{(h)}}{v_{n,m}^{(h)}}-p\frac{v_{n,m+1}^{(h)}}{v_{n,m}^{(h)}}, \label{dGD:MTa} \\
 &p+u_{n,m}^{(h+1)}-u_{n+1,m}^{(h)}=p\frac{v_{n+1,m}^{(h)}}{v_{n,m}^{(h)}}, \quad q+u_{n,m}^{(h+1)}-u_{n,m+1}^{(h)}=q\frac{v_{n,m+1}^{(h)}}{v_{n,m}^{(h)}}, \label{dGD:MTb}
\end{align}
\ese
which actually play the role of discrete Miura-type transforms, though the transforms here are nonlocal in both directions.
Adding the three equations in \eqref{dGD:MT} up, the additive potential $u$ is eliminated, and it turns out that we derive a closed-form equation for only $v$,
which is in the form of
\begin{align}\label{dGD:v}
 p\left(\frac{v_{n+1,m+1}^{(h)}}{v_{n,m+1}^{(h)}}-\frac{v_{n+1,m}^{(h+1)}}{v_{n,m}^{(h+1)}}\right)
 =q\left(\frac{v_{n+1,m+1}^{(h)}}{v_{n+1,m}^{(h)}}-\frac{v_{n,m+1}^{(h+1)}}{v_{n,m}^{(h+1)}}\right).
\end{align}
Notice that $v$ satisfies the periodicity \eqref{dGD:uvPeriod}.
Equation \eqref{dGD:v} together with $v_{n,m}^{(h+\cN)}=v_{n,m}^{(h)}$ for $h=0,1,\cdots,\cN-1$ forms the $(\alpha,\beta)=(0,0)$ case of \eqref{dFX:v}.

Next, we prove that the component
\begin{align}\label{phi}
 \phi_{n,m}^{(h)}(k)\doteq\Phi_{n,m}^{(h)}(0;k)
\end{align}
is the wave function for the Lax pairs of equations \eqref{dGD:u} and \eqref{dGD:v}, respectively.
This can be realised by constructing the Lax pairs by starting from \eqref{Linear}.
Remind ourselves that $\bU$ obeys the periodic condition $\bU_{n,m}^{(h+\cN)}=\bU_{n,m}^{(h)}$
and the plane wave factor $\rho$ satisfies $\rho_{n,m}^{(h+\cN)}(k)=k^\cN\rho_{n,m}^{(h)}(k)$.
The relation \eqref{Linear} yields the quasi-periodic condition
\begin{align*}
 \Phi_{n,m}^{(h+\cN)}(k)=k^\cN\Phi_{n,m}^{(h)}(k), \quad \hbox{which leads to} \quad \phi_{n,m}^{(h+\cN)}(k)=k^\cN\phi_{n,m}^{(h)}(k),
\end{align*}
because $\phi$ is a component of $\Phi$. Shifting the wave function $\Phi$ with respect to $n$ by one unit, we have
\begin{align*}
 \Phi_{n+1,m}^{(h)}(k)&=\left(1-\bU_{n+1,m}^{(h)}\bOa\right)\bc(k)\rho_{n+1,m}^{(h)}(k)=\left(1-\bU_{n+1,m}^{(h)}\bOa\right)(p+k)\bc(k)\rho_{n,m}^{(h)}(k) \\
 &=\left(1-\bU_{n+1,m}^{(h)}\bOa\right)(p+\bLd)\bc(k)\rho_{n,m}^{(h)}(k)=(p+\bLd)\bc(k)\rho_{n,m}^{(h)}(k)-\bU_{n+1,m}^{(h)}\bOa(p+\bLd)\bc(k)\rho_{n,m}^{(h)}(k) \\
 &=(p+\bLd)\bc(k)\rho_{n,m}^{(h)}(k)-\bU_{n+1,m}^{(h)}\bO\bc(k)\rho_{n,m}^{(h)}(k)-\bU_{n+1,m}^{(h)}(p+\tbLd)\bOa\bc(k)\rho_{n,m}^{(h)}(k) \\
 &=(p+\bLd)\bc(k)\rho_{n,m}^{(h)}(k)-\bU_{n+1,m}^{(h)}\bO\bc(k)\rho_{n,m}^{(h)}(k)-\left[(p+\bLd)\bU_{n,m}^{(h)}-\bU_{n+1,m}^{(h)}\bO\bU_{n,m}^{(h)}\right]\bOa\bc(k)\rho_{n,m}^{(h)}(k) \\
 &=(p+\bLd)\left(1-\bU_{n,m}^{(h)}\bOa\right)\bc(k)\rho_{n,m}^{(h)}(k)-\bU_{n+1,m}^{(h)}\bO\left(1-\bU_{n,m}^{(h)}\bOa\right)\bc(k)\rho_{n,m}^{(h)}(k),
\end{align*}
which implies that the wave function $\Phi$ satisfies a dynamical relation
\bse\label{dGD:PhiDyn}
\begin{align}
 \Phi_{n+1,m}^{(h)}=(p+\bLd)\Phi_{n,m}^{(h)}-\bU_{n+1,m}^{(h)}\bO\Phi_{n,m}^{(h)}, \label{dGD:PhiDyna}
\end{align}
where for the second last equality we have made use of equation \eqref{Linear}. Likewise, we have the following two relations for the $m$ and $h$ directions:
\begin{align}
 &\Phi_{n,m+1}^{(h)}=(q+\bLd)\Phi_{n,m}^{(h)}-\bU_{n,m+1}^{(h)}\bO\Phi_{n,m}^{(h)}, \label{dGD:PhiDynb} \\
 &\Phi_{n,m}^{(h+1)}=\bLd\Phi_{n,m}^{(h)}-\bU_{n,m}^{(h+1)}\bO\Phi_{n,m}^{(h)}. \label{dGD:PhiDync}
\end{align}
\ese
Subtracting \eqref{dGD:PhiDync} from \eqref{dGD:PhiDyna} and \eqref{dGD:PhiDynb}, respectively,
we are able to eliminate $\bLd$ and obtain the following dynamical relations:
\bse
\begin{align*}
 &\Phi_{n+1,m}^{(h)}=\left(p+\bU_{n,m}^{(h+1)}\bO-\bU_{n+1,m}^{(h)}\bO\right)\Phi_{n,m}^{(h)}+\Phi_{n,m}^{(h+1)}, \\
 &\Phi_{n,m+1}^{(h)}=\left(q+\bU_{n,m}^{(h+1)}\bO-\bU_{n,m+1}^{(h)}\bO\right)\Phi_{n,m}^{(h)}+\Phi_{n,m}^{(h+1)},
\end{align*}
\ese
These two equations will lead to the Lax pair for equation \eqref{dGD:u}. In fact, by taking the $0$-component of the two equations, we immediately get
\begin{align}\label{dGD:Lax}
 \phi_{n+1,m}^{(h)}=\left(p+u_{n,m}^{(h+1)}-u_{n+1,m}^{(h)}\right)\phi_{n,m}^{(h)}+\phi_{n,m}^{(h+1)}, \quad
 \phi_{n,m+1}^{(h)}=\left(q+u_{n,m}^{(h+1)}-u_{n,m+1}^{(h)}\right)\phi_{n,m}^{(h)}+\phi_{n,m}^{(h+1)},
\end{align}
which together with the quasi-periodic condition $\phi_{n,m}^{(h+\cN)}=k^\cN\phi_{n,m}^{(h)}$ forms the $\mathbb{Z}_\cN$ graded Lax pair for \eqref{dGD:u},
for $h=0,1,\cdots,\cN-1$. Replacing $u$ by $v$ with the help of the discrete Miura transforms given in \eqref{dGD:MT},
the Lax pair for equation \eqref{dGD:v}, namely
\begin{align*}
 \phi_{n+1,m}^{(h)}=p\frac{v_{n+1,m}^{(h)}}{v_{n,m}^{(h)}}\phi_{n,m}^{(h)}+\phi_{n,m}^{(h+1)}, \quad
 \phi_{n,m+1}^{(h)}=q\frac{v_{n,m+1}^{(h)}}{v_{n,m}^{(h)}}\phi_{n,m}^{(h)}+\phi_{n,m}^{(h+1)}, \quad
 \phi_{n,m}^{(h+\cN)}=k^\cN\phi_{n,m}^{(h)},
\end{align*}
in which $h=0,1,\cdots,\cN-1$, is obtained.

We have proven the $(\alpha,\beta)=0$ case of theorem \ref{T:DL}, and now consider the case for generic $\alpha$ and $\beta$.
Comparing the plane wave factors \eqref{dGD:PWF} and \eqref{dFX:PWF},
we can observe that the latter is a parameter deformation of the former, by introducing a change of variables $(n,m,h)\mapsto(n,m,r)$ as follows:
\begin{align*}
 n\doteq n, \quad m\doteq m, \quad h\doteq r+\alpha n+\beta m,
\end{align*}
where $\alpha,\beta\in\{0,1,\cdots,\cN-1\}$. The parameter deformation results in the dynamical relations
$\rho_{n+1,m}^{(r)}=\rho_{n+1,m}^{(h+\alpha)}$, $\rho_{n,m+1}^{(r)}=\rho_{n,m+1}^{(h+\beta)}$,
and $\rho_{n,m}^{(r+1)}=\rho_{n,m}^{(h+1)}$ for $\rho$ as well as the same relations for $\sigma$.
Since the dynamics of $\bC$ explicitly relies on the product of $\rho$ and $\sigma$, one can verify that the infinite matrix $\bC$ satisfies
\begin{align}\label{dFX:CTransform}
 \bC_{n+1,m}^{(r)}=\bC_{n+1,m}^{(h+\alpha)}, \quad
 \bC_{n,m+1}^{(r)}=\bC_{n,m+1}^{(h+\beta)} \quad \hbox{and} \quad
 \bC_{n,m}^{(r+1)}=\bC_{n,m}^{(h+1)},
\end{align}
as they follow from \eqref{C} and \eqref{dFX:C} directly.
Furthermore, the infinite matrix $\bU$ obeys relations $\bU_{n+1,m}^{(r)}=\bU_{n+1,m}^{(h+\alpha)}$,
$\bU_{n,m+1}^{(r)}=\bU_{n,m+1}^{(h+\beta)}$ and $\bU_{n,m}^{(r+1)}=\bU_{n,m}^{(h+1)}$,
as the dynamics of $\bU$ only depends on $\bC$ through the equation $\bU=(1-\bU\bOa)\bC$, cf. \eqref{Nonlinear},
which consequently together with the $\rho$ relations result in the same deformation on the wave function $\Phi$, namely
$\Phi_{n+1,m}^{(r)}=\Phi_{n+1,m}^{(h+\alpha)}$, $\Phi_{n,m+1}^{(r)}=\Phi_{n,m+1}^{(h+\beta)}$,
and $\Phi_{n,m}^{(r+1)}=\Phi_{n,m}^{(h+1)}$.
Noticing that the additive potential $u$ and the quotient potential $v$, and the wave function $\phi$
are entries and the $0$th-component of $\bU$ and $\Phi$, respectively, see \eqref{uv} and \eqref{phi},
we conclude that these variables obey exactly the same deformation relations.
And thus, by applying the deformation to nonlinear equations \eqref{dGD:u} and \eqref{dGD:v}, the difference transform \eqref{dGD:MT}, and the Lax pair \eqref{dGD:Lax},
one obtains the \ac{FX} discrete models \eqref{dFX:u} and \eqref{dFX:v}, the relevant \ac{BT} \eqref{dFX:MT}, as well as the $\mathbb{Z}_\cN$ graded Lax pair \eqref{dFX:Lax}.
To put it in another way, based on the linear integral equation
\begin{align}
 \Phi_{n,m}^{(r)}(k)
 +\sum_{j\in J}\int_{\Gamma_j}\rd\ld_j(l)\frac{\rho_{n,m}^{(r)}(k)\sigma_{n,m}^{(r)}(\oa^jl)}{k-\oa^jl}\Phi_{n,m}^{(r)}(l)
 =\rho_{n,m}^{(r)}(k)\bc(k),
\end{align}
we have proven that $u=U(0,0)$ and $v=1+U(0,-1)$ solve the nonlinear equations \eqref{dFX:u} and \eqref{dFX:v},
and $\phi(k)=\Phi(0,k)$ solves the associated Lax pair \eqref{dFX:Lax}, which is indeed the result of theorem \ref{T:DL}.

\section{Proof of theorem \ref{T:BL}}\label{S:BL}

We still start with the $(\alpha,\beta)=(0,0)$ case. The idea is to establish the link between dynamics of the tau functions and entries of the infinite matrix $\bU$.
By making use of the infinite matrix relations found in section \ref{S:DL},
a closed-form multicomponent system expressed by only the tau functions, namely the bilinear form of the \ac{FX} nonlinear discrete systems is therefore obtained.

The tau function in this case is defined as $\tau_{n,m}^{(h)}=\det\left(1+\bOa\bC_{n,m}^{(h)}\right)$,
in which the determinant of infinite matrix should be thought of as the expansion of
\begin{align}\label{dGD:tauFunction}
 \exp\left(\tr\ln\left(1+\bOa\bC_{n,m}^{(h)}\right)\right), \quad \hbox{namely} \quad
 \tau_{n,m}^{(h)}=\sum_{i'=0}^{\infty}\frac{1}{i'!}\left(\sum_{i=0}^\infty\frac{(-1)^{i}}{i+1}\tr\left(\bOa\bC_{n,m}^{(h)}\right)^{i+1}\right)^{i'},
\end{align}
as the determinant of an infinite matrix is defined thorough $\ln\det(\bA)=\tr\ln(\bA)$.
A comment here is that the infinite summation does not lead to divergence,
because the infinite matrix $\bOa$ depends on $\bO$ and the trace operation guarantees the convergence.
From the definition of the tau function, one can immediately deduce that the tau function satisfies the periodic condition
\begin{align}\label{dGD:tauPeriod}
 \tau_{n,m}^{(h+\cN)}=\tau_{n,m}^{(h)}.
\end{align}
This is because the dynamics of $\tau$ is completely governed by $\bC$ according the definition \eqref{dGD:tauFunction},
and therefore the tau function inherits the periodicity of $\bC$, cf. \eqref{dGD:Period}.

Next, we consider the dynamics of the tau function. For future convenience, we introduce a new quantity
\begin{align*}
 V_{n,m}^{(h)}(a)\doteq 1+\left(\bU_{n,m}^{(h)}(a+\tbLd)^{-1}\right)(0,0),
\end{align*}
where $(\cdot)(0,0)$ denotes taking the $(0,0)$-entry of the infinite matrix in the bracket, and $a$ is a parameter.
Now we establish the link between the dynamics of the tau function and the entries of the infinite matrix $\bU$.
Direct calculation shows that
\begin{align*}
 \tau_{n+1,m}^{(h)}&=\det\left(1+\bOa\bC_{n+1,m}^{(h)}\right)=\det\left(1+\bOa(p+\bLd)\bC_{n,m}^{(h)}(p+\tbLd)^{-1}\right) \\
 &=\det\left(1+\left(\bO+(p+\tbLd)\bOa\right)\bC_{n,m}^{(h)}(p+\tbLd)^{-1}\right)=\det\left(1+\bOa\bC_{n,m}^{(h)}+(p+\tbLd)^{-1}\bO\bC_{n,m}^{(h)}\right) \\
 &=\det\left(1+\bOa\bC_{n,m}^{(h)}\right)\det\left(\left(1+\bOa\bC_{n,m}^{(h)}\right)^{-1}(p+\tbLd)^{-1}\bO\bC_{n,m}^{(h)}\right) \\
 &=\tau_{n,m}^{(h)}\left(1+\left(\bC_{n,m}^{(h)}\left(1+\bOa\bC_{n,m}^{(h)}\right)^{-1}(p+\tbLd)^{-1}\right)(0,0)\right)
 =\tau_{n,m}^{(h)}\left(1+\left(\bU_{n,m}^{(h)}(p+\tbLd)^{-1}\right)(0,0)\right),
\end{align*}
where the Weinstein--Aronszajn formula is used for the second equality. This gives rise to
\begin{align}\label{dGD:tauDyn}
 \frac{\tau_{n+1,m}^{(h)}}{\tau_{n,m}^{(h)}}=V_{n,m}^{(h)}(p), \quad \hbox{and similarly}, \quad
 \frac{\tau_{n,m+1}^{(h)}}{\tau_{n,m}^{(h)}}=V_{n,m}^{(h)}(q) \quad \hbox{and} \quad \frac{\tau_{n,m}^{(h+1)}}{\tau_{n,m}^{(h)}}=v_{n,m}^{(h)}.
\end{align}
These equalities form the fundamental relations to construct the bilinear form of the \ac{FX} discrete equations.
Multiplying equation \eqref{dGD:UDyna} by $(a+\tbLd)^{-1}$, the following equation is obtained:
\begin{align*}
 \bU_{n+1,m}^{(h)}\frac{p+\tbLd}{a+\tbLd}=(p-a)\bU_{n+1,m}^{(h)}(a+\tbLd)^{-1}+\bU_{n+1,m}^{(h)}
 =(p+\bLd)\bU_{n,m}^{(h)}(a+\tbLd)^{-1}-\bU_{n+1,m}^{(h)}\bO\bU_{n,m}^{(h)}(a+\tbLd)^{-1}.
\end{align*}
Taking the $(0,0)$-entry of this matrix equation, we can derive
\begin{align*}
 u_{n+1,m}^{(h)}V_{n,m}^{(h)}(a)+a=pV_{n,m}^{(h)}(a)+\left(\bLd\bU(a+\tbLd)^{-1}\right)(0,0).
\end{align*}
Similarly, we have also
\begin{align*}
 u_{n,m+1}^{(h)}V_{n,m}^{(h)}(a)+a=qV_{n,m}^{(h)}(a)+\left(\bLd\bU(a+\tbLd)^{-1}\right)(0,0)
\end{align*}
for the $m$-direction. Subtracting the above two equations, the entry $\left(\bLd\bU(a+\tbLd)^{-1}\right)(0,0)$ is eliminated,
and thus, we end up with a difference relation between $u$ and $V(a)$, which is given by
\begin{align*}
 p-q+u_{n,m+1}^{(h)}-u_{n+1,m}^{(h)}=(p-a)\frac{V_{n+1,m}^{(h)}(a)}{V_{n,m}^{(h)}(a)}-(q-a)\frac{V_{n,m+1}^{(h)}(a)}{V_{n,m}^{(h)}(a)}.
\end{align*}
By setting $a=q$ and taking the first two equations in \eqref{dGD:tauDyn} into consideration,
we can replace the $V$-variable by the tau function, and consequently obtain
\bse\label{dGD:BLT}
\begin{align}
 p-q+u_{n,m+1}^{(h)}-u_{n+1,m}^{(h)}=(p-q)\frac{\tau_{n,m}^{(h)}\tau_{n+1,m+1}^{(h)}}{\tau_{n,m+1}^{(h)}\tau_{n+1,m}^{(h)}}.
\end{align}
Meanwhile, the third equation in \eqref{dGD:tauDyn} establishes the connection between the quotient potential $v$ and the tau function,
which together with equations \eqref{dGD:MTb} results in the following equations:
\begin{align}
 p+u_{n,m}^{(h+1)}-u_{n+1,m}^{(h)}=p\frac{\tau_{n,m}^{(h)}\tau_{n+1,m}^{(h+1)}}{\tau_{n,m}^{(h+1)}\tau_{n+1,m}^{(h)}} \quad \hbox{and} \quad
 q+u_{n,m}^{(h+1)}-u_{n,m+1}^{(h)}=q\frac{\tau_{n,m}^{(h)}\tau_{n,m+1}^{(h+1)}}{\tau_{n,m}^{(h+1)}\tau_{n,m+1}^{(h)}}
\end{align}
\ese
Eliminating $u$ in \eqref{dGD:BLT}, a bilinear equation arises, which takes the form of
\begin{align}\label{dGD:tau}
 p\left(\tau_{n,m}^{(h+1)}\tau_{n+1,m+1}^{(h)}-\tau_{n,m+1}^{(h)}\tau_{n+1,m}^{(h+1)}\right)
 =q\left(\tau_{n,m}^{(h+1)}\tau_{n+1,m+1}^{(h)}-\tau_{n+1,m}^{(h)}\tau_{n,m+1}^{(h+1)}\right).
\end{align}
Equation \eqref{dGD:tau} subject to the periodic relation \eqref{dGD:tauPeriod} for $h=0,1,\cdots,\cN-1$ forms an $\cN$-component coupled discrete bilinear system,
which is the bilinear version of the \ac{FX} models when $(\alpha,\beta)=(0,0)$.

We can now apply the parameter deformation to the bilinear system \eqref{dGD:tau}.
Notice that in \eqref{dFX:tauFunction} the dynamics of the tau function $\tau$ is completely determined by $\bC$, we thus have a similar change of variables, namely
\begin{align}\label{dFX:tauTransform}
 \tau_{n+1,m}^{(r)}=\tau_{n+1,m}^{(h+\alpha)}, \quad
 \tau_{n,m+1}^{(r)}=\tau_{n,m+1}^{(h+\beta)} \quad \hbox{and} \quad
 \tau_{n,m}^{(r+1)}=\tau_{n,m}^{(h+1)}.
\end{align}
The application of equation \eqref{dFX:tauTransform} to \eqref{dGD:tau} gives us the bilinear equation \eqref{dFX:tau}.
Simultaneously, the tau function \eqref{dGD:tauFunction} turns out to be \eqref{dFX:tauFunction}.
Moreover, we note that the acting the same change of variables on \eqref{dGD:BLT} and \eqref{dGD:tauDyn} also provides us with \eqref{dFX:BLT},
which together with the periodicity condition of the tau function, namely $\tau_{n,m}^{(r+\cN)}=\tau_{n,m}^{(r)}$, yield the constraints \eqref{dFX:Identity}.
Theorem \ref{T:BL} is proven.

\section{Concluding remarks}\label{S:Concl}

In this paper, theorems \ref{T:DL} and \ref{T:BL} provide solutions involving an integral with respect to the spectral variable to
the \ac{FX} models \eqref{dFX:u} and \eqref{dFX:v}, and the bilinear equation \eqref{dFX:tau}, respectively.
Our main motivation is to explain the solution structure of the integrable difference equations proposed by Fordy and Xenitidis.
In fact, in the solution structure, the spectral variable sits in a continuous space;
by specifying the measure associated with the linear integral, special classes of solutions to these discrete systems will naturally arise.
Particularly, reduction of the measure to a discrete space, i.e. introducing a finite number of singularities in the measure,
will lead to the degeneracy of the linear integral, providing determinantal (namely soliton-type) solutions.

We take equation \eqref{dFX:9pt} as an example. In this case $\cN=3$, $\alpha=\beta=1$, and the linear integral equation \eqref{dFX:Integral} becomes
\begin{align}\label{SPIntegral}
 \phi_{n,m}^{(r)}(k)
 +\int_{\Gamma_1}\rd\ld_1(l)\frac{\rho_{n,m}^{(r)}(k)\sigma_{n,m}^{(r)}(\oa l)}{k-\oa l}\phi_{n,m}^{(r)}(l)
 +\int_{\Gamma_2}\rd\ld_2(l)\frac{\rho_{n,m}^{(r)}(k)\sigma_{n,m}^{(r)}(\oa^2 l)}{k-\oa^2 l}\phi_{n,m}^{(r)}(l)
 =\rho_{n,m}^{(r)}(k),
\end{align}
where the plane wave factors satisfy
\begin{align*}
 \rho_{n,m}^{(r)}(k)=(k(p+k))^{n}(k(q+k))^{m}k^r \quad \hbox{and} \quad
 \sigma_{n,m}^{(r)}(k')=(k'(p+k'))^{-n}(k'(q+k'))^{-m}k'^{-r},
\end{align*}
and $\omega=\exp\{2\pi\mathfrak{i}/3\}$. Accordingly, by following \eqref{dFX:uv}, we have
\begin{align}\label{SPPotential}
 v_{n,m}^{(r)}=1+\int_{\Gamma_1}\rd\ld_1(k)\phi_{n,m}^{(r)}(k)\sigma_{n,m}^{(r)}(\oa k)\oa^{-1}k^{-1}
 +\int_{\Gamma_2}\rd\ld_2(k)\phi_{n,m}^{(r)}(k)\sigma_{n,m}^{(r)}(\oa^2k)\oa^{-2}k^{-1}.
\end{align}
Now we take very special measures
\begin{align*}
 \rd\ld_1(l)=\frac{1}{2\pi\mathfrak{i}}\frac{1}{k-k_1}\rd l \quad \hbox{and} \quad \rd\ld_2(l)=\frac{1}{2\pi\mathfrak{i}}\frac{1}{k-k_2}\rd l,
\end{align*}
and require that the contours $\Gamma_1$ and $\Gamma_2$ contain $k_1$ and $k_2$ as their only singularities, respectively.
As a result, by the residue theorem equation \eqref{SPIntegral} yields
\begin{align*}
 \phi_{n,m}^{(r)}(k)+\frac{\rho_{n,m}^{(r)}(k)\sigma_{n,m}^{(r)}(\oa k_1)}{k-\oa k_1}\phi_{n,m}^{(r)}(k_1)
 +\frac{\rho_{n,m}^{(r)}(k)\sigma_{n,m}^{(r)}(\oa k_2)}{k-\oa k_2}\phi_{n,m}^{(r)}(k_2)=\rho_k,
\end{align*}
which in turn implies
\begin{align*}
\begin{pmatrix}
 1+\frac{\rho_{n,m}^{(r)}(k_1)\sigma_{n,m}^{(r)}(\oa k_1)}{k_1-\oa k_1} & \frac{\rho_{n,m}^{(r)}(k_1)\sigma_{n,m}^{(r)}(\oa k_2)}{k_1-\oa k_2} \\
 \frac{\rho_{n,m}^{(r)}(k_2)\sigma_{n,m}^{(r)}(\oa k_1)}{k_2-\oa k_1} & 1+\frac{\rho_{n,m}^{(r)}(k_2)\sigma_{n,m}^{(r)}(\oa k_2)}{k_2-\oa k_2}
\end{pmatrix}
\begin{pmatrix}
 \phi_{n,m}^{(r)}(k_1) \\
 \phi_{n,m}^{(r)}(k_2)
\end{pmatrix}
=
\begin{pmatrix}
 \rho_{n,m}^{(r)}(k_1) \\
 \rho_{n,m}^{(r)}(k_2)
\end{pmatrix}
\end{align*}
if $k$ is set to be $k_1$ and $k_2$, respectively. Simultaneously, the special measure also give rise to a reduction of the potential $v$ given by
\begin{align*}
v_{n,m}^{(r)}={}&1+\phi_{n,m}^{(r)}(k_1)\sigma_{n,m}^{(r)}(\oa k_1)\oa^{-1}k_1^{-1}+\phi_{n,m}^{(r)}(k_2)\sigma_{n,m}^{(r)}(\oa^2k_2)\oa^{-2}k_2^{-1} \\
={}&
\left(\sigma_{n,m}^{(r)}(\oa k_1),\sigma_{n,m}^{(r)}(\oa^2k_2)\right)
\begin{pmatrix}
 \omega k_1 & 0 \\
 0 & \omega^2 k_2
\end{pmatrix}^{-1}
\begin{pmatrix}
 \phi_{n,m}^{(r)}(k_1) \\
 \phi_{n,m}^{(r)}(k_2)
\end{pmatrix}.
\end{align*}
And thus, we obtain
\begin{align*}
 v_{n,m}^{(r)}=1+\left(\sigma_{n,m}^{(r)}(\oa k_1),\sigma_{n,m}^{(r)}(\oa^2k_2)\right)
\begin{pmatrix}
 \omega k_1 & 0 \\
 0 & \omega^2 k_2
\end{pmatrix}^{-1}
\begin{pmatrix}
 1+\frac{\rho_{n,m}^{(r)}(k_1)\sigma_{n,m}^{(r)}(\oa k_1)}{k_1-\oa k_1} & \frac{\rho_{n,m}^{(r)}(k_1)\sigma_{n,m}^{(r)}(\oa k_2)}{k_1-\oa k_2} \\
 \frac{\rho_{n,m}^{(r)}(k_2)\sigma_{n,m}^{(r)}(\oa k_1)}{k_2-\oa k_1} & 1+\frac{\rho_{n,m}^{(r)}(k_2)\sigma_{n,m}^{(r)}(\oa k_2)}{k_2-\oa k_2}
\end{pmatrix}^{-1}
\begin{pmatrix}
 \rho_{n,m}^{(r)}(k_1) \\
 \rho_{n,m}^{(r)}(k_2)
\end{pmatrix}.
\end{align*}
Although the above formula contain two parameters $k_1$ and $k_2$,
we note that for any $r=0,1,2$ the obtained $v$ plays the role of the one-soliton solution to the \ac{FX} equation.
This is because the two parameters really follow from the two distinct measures and they do not govern soliton interaction;
in other words, the result obtained from the \ac{DL} method in a sense generalises the traditional one-soliton solution.

Another remark is that for fixed $\cN$, all \ac{FX} discrete equations share the same form of linear integral equation \eqref{dFX:Integral}
but with different plane wave factors \eqref{dFX:PWF}. In other words, from the viewpoint of solution structure,
the \ac{FX} systems are distinguished by the linear dispersion, while the nonlinear structures (i.e. the behaviours of the spectral variable in difference equations) are the same.
As the $(\alpha,\beta)=(0,0)$ case describes the structure of the standard discrete \ac{GD} hierarchy,
the novel equations in the \ac{FX} in a sense can also be thought of as the equations governed by deformed discrete flows in the discrete \ac{GD} hierarchy.

We only consider the coprime case of the \ac{FX} classification in this case.
It is not yet clear whether discrete equations in the non-coprime case can also be solved from the \ac{DL} approach,
but it is quite certain that solution structure of those equations is entirely different from that of the discrete \ac{GD} hierarchy at least.
This remains a problem for our future work.

\section*{Acknowledgments}
This project was sponsored by the National Natural Science Foundation of China (grant no. 11901198) and by Shanghai Pujiang Program (grant no. 19PJ1403200).
The author was also partially supported by the Science and Technology Commission of Shanghai Municipality (grant no. 18dz2271000),
and he thanks Frank Nijhoff for comments on a draft version of this manuscript.

\renewcommand{\bibname}{References}
\bibliography{References}
\bibliographystyle{amsplain}

\end{document}